# Efficacy of Object-Based Passwords for User Authentication

Sufian Hameed, Lamak Qaizar, Shankar Khatri
CS Department, FAST-NUCES, Karachi Campus, Pakistan.
sufian.hameed@nu.edu.pk, k092031@nu.edu.pk, k092104@nu.edu.pk

*Abstract*—Traditional text-based password schemes are inherently weak. Users tend to choose passwords that are easy to remember, making them susceptible to various attacks that have matured over the years. *ObPwd* [5] has tried to address these issues by converting user-selected digital objects to high-entropy text passwords for user authentication.

In this paper, we extend the *ObPwd* scheme with a new object based password scheme that performs majority of the computation at the server side. This paper essentially discusses two frameworks for object password schemes, an object hash-based scheme (where the client machine computes the hash of the object to be used as text password) and an object-based scheme (where the object is directly transmitted to the server as password). We also evaluate the performance of both the object password schemes against conventional text-based password schemes using prototypes of each of the frameworks. Implications with respect to ease of use, sharing and security are also discussed.

## I. INTRODUCTION

Users pay insufficient attention to wisely choose a password. This tendency has been exploited using simple attacks like password guessing to more specialized methods such as dictionary attacks [8]. The loopholes of text-based passwords are well documented [10], [15], [16], [19]. It is also difficult for users to generate and memorize strong or high-entropy passwords. Further, these strong passwords are generally usable only if they are frequently used. Passwords for rarely-used services are hard to reproduce at a later point in time.

In *ObPwd* [5], Mannan et.al. attempted to address these issues by instructing the users to use digital objects as passwords. *ObPwd* converts the user-selected digital objects to high-entropy text passwords for user authentication. These digital objects may include images, videos, audio, documents and e-books in different formats. Essentially such passwords are not only easy to remember through semantic clues, but are also impossible to crack in linear time [2] when accompanied by a secure hash function coupled with unique media objects.

In this paper, we extend *ObPwd* with a new object based password scheme that performs majority of the computation at the server side. This paper essentially presents two frameworks that leverage digital objects as password; 1) an object hash-based scheme and 2) an object-based scheme.

*Object-based scheme:* In this scheme the digital object is directly transmitted to the server where all computation takes place. This incurs a network overhead but frees the client of hash computation. This approach is beneficial for the low end client machines with insufficient processing power.

*Object hash-based scheme:* This scheme involves the computation of the object hash on the client system. The hash is then transmitted to the server for verification and login. The basic idea of object hash-based scheme is essentially the same as *ObPwd* [5]. However, a stronger hash function is used in this scheme, i.e. SHA-256, instead of SHA-1 used in *ObPwd*. Furthermore, *ObPwd* uses *PwdHash* [18] for converting hash values into a 12 characters long alphanumeric password and restricts the object size between 30 and 100,000 bytes. The object hash-based scheme is free from any such restrictions.

We have developed three prototype frameworks which follow the client-server architecture for text, object-hash and object based password schemes. Client-side scripts are written in JavaScript whereas *PHP* scripts are deployed on the server along with a *MySQL* database. The database stores user account details; *user-IDs*, password hashes and 128-bit salts in particular. Using the prototype implementations of the frameworks we evaluate the performance of both the object password schemes against conventional text-based password scheme.

The rest of the paper is organized as follows. §II describes the state of the art. §III discusses the design and prototype of different frameworks. In §IV we discuss the usage flow of different schemes in the prototype deployed on the web server. In §V we demonstrate the effectiveness and complexity of object based password schemes using system evaluations. In §VI we discuss how the different password schemes offer certain benefits over one another. Finally we conclude the paper in §VII.

## II. RELATED WORK

Numerous efforts have been made over the years by the scientific community to strengthen passwords and to enhance their usability. A few noteworthy efforts include [4], [6], [7], [9], [11], [12], [20], [21]. In this section we have primarily focused on the most relevant object based password scheme.

The idea of using digital objects as password was first realized in *ObPwd* by Mannan et.al. [5], [14]. It allows users to generate passwords from digital content that may range from a personal collection of photographs to static content from the web. The Internet Archive (www.archive.org) and Google Books (books.google.com) were recognized as good object selection pools. A prototype browser based plug-in was developed that computed object hashes on the client machine.

The resulting hash string could be copied to the clipboard and used as a *text-based password* by the user. It allows the hash to be recorded (because of its textual nature) on any medium, and also allowed the hash to be recomputed, provided the same object could be reproduced by the user.

In this paper, we extend *ObPwd* scheme with a new object based password scheme that offloads majority of computation at the server side. This approach is beneficial for the low end client machines with insufficient processing power.

## III. FRAMEWORK DESIGN AND PROTOTYPE

We have developed prototypes of three different password schemes; 1) text-based, 2) object hash-based and 3) object-based. The framework design and prototypes are discussed in the following subsections.

### A. Text-Based Password Scheme

Text-based passwords have been used since ancient times to allow (or disallow) a person or group to enter an area. They have since been adopted by information system designers to serve the same purpose of authentication and gaining access to a resource.

Fig. 1 shows how a standard text-based password scheme is realized to authenticate any user over the web. A user supplies their credentials (a *user-ID* and password) through a web form to authenticate themselves to an information system and to gain access to its online services. The *user-ID* and password are delivered to the authenticating web-server, which then decides whether the supplied credentials are correct or incorrect.

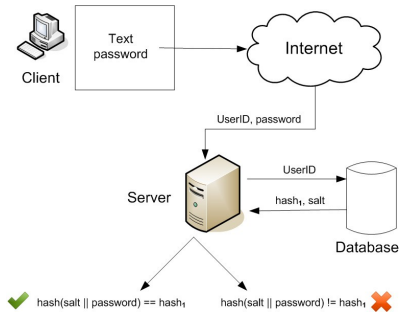

Fig. 1. Text-based password scheme.

While spoofing may be mitigated using *SSL*, text-based passwords still leave the door open for many other attacks. Most of these attacks arise because of the user's choice of passwords. Text-based passwords actually have to be *remembered*, imposing a memory load on the user that they would otherwise wish to avoid. Enterprises place various restrictions on the user's choice of password (length, numeric or special characters) so that they may choose stronger, more secure passwords. This technique is not very effective however, as users still look for the easiest password they can construct given the restrictions. Meanwhile attackers can compute special tables with most common user passwords and use them to break into user accounts [17].

For the prototype of text-based scheme, the server maintains a list of *user-IDs* and their respective password hashes in a database. Passwords are not stored as plain-text to account for any possibility of the database being compromised, in which case the attacker would have access to every user account on the system. An attacker may still however employ rainbow tables to crack these password hashes, and to mitigate such an attack, a key derivation function using salts is employed. A salt is a randomly generated string that is generated with each new user account that is created.

Upon an authentication request, the server fetches the corresponding password hash and salt stored against the *user-ID*. It then concatenates the salt with the password, computes the hash and validates it against the hash fetched from the database. The cryptographic hash function utilized here is *SHA-256*, from the *SHA-2* family.

### B. Object Hash-Based Password Scheme

The object hash-based password scheme employs media objects as user's passwords and performs majority of the computation on the client using client-side scripts.

As it can be seen from Fig. 2, the server side functionality is consistent with that of the text-based password scheme. The server receives the *user-ID* and a text-based version of the user password (hash of the media object), fetches the password hash and salt from the database and evaluates whether the received credentials are valid.

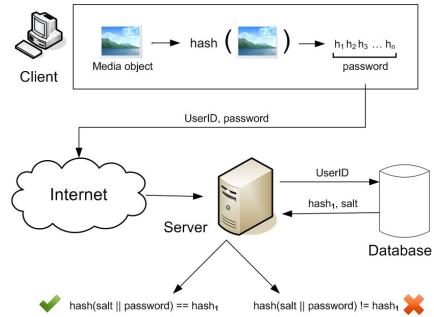

Fig. 2. Object hash-based password scheme.

This scheme allows the user to maintain a text-based version of their chosen media object on a non-digital medium, in case their media object is unavailable at any point in time. This would be highly useful if and when the media object were to become corrupted, and also when the user would be trying to gain access to a service from a system not of their own, and the media object would either be unavailable or could possibly be compromised if loaded on a 3rd party computer.

The server-side implementation is identical to that of the text-based password scheme. The server receives the *user-ID* and a text-based version of the media object as opposed to the password in plaintext from the text-based scheme. However, since both of them are strings (character sequences), the subsequent computation is the same.

The client on the other hand requires some additional computation. A hash value (using *SHA-256*) is computed for the media object using a client-side script written in JavaScript.

The basic idea of this scheme is essentially the same as *ObPwd* [5]. However, we use a stronger hash function i.e. *SHA-256*, instead of *SHA-1* used in *ObPwd*. Further, *ObPwd* use *PwdHash* [18] for reducing the hash values into a 12 characters long alphanumeric password and restricts the object size between 30 and 100000 bytes. Our solution does not apply any such restrictions.

## C. Object-Based Password Scheme

We extend the *ObPwd* scheme with a new object based password scheme that performs majority of the computation at the server side.

This scheme extends the previous object hash-based scheme in such a way that majority of the computation is offloaded at the server side. In this scheme the media object selected by the user is directly sent to the server without any processing at the client side.

Fig. 3 highlights the differences of the object-based scheme with the text-based and the object hash-based schemes. When the digital object is transmitted, the server takes over the responsibility of computing the initial hash of the object. Once the text-based representation of the object is produced (i.e. the hash of the object), the server proceeds to validate authenticity of user based on the produced string.

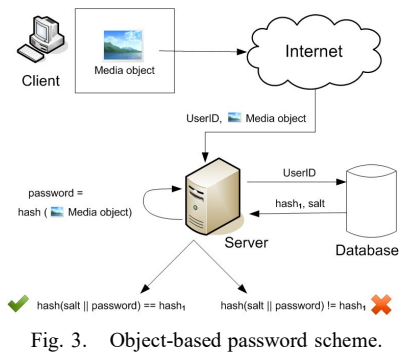

Fig. 3. Object-based password scheme.

In this scheme the client is not required to perform any processing, and it is only responsible for uploading the media object to the server. Once the object is uploaded, a server-side script written in *PHP* computes the hash of the object using *SHA-256*, the produced hash we refer to as the password. As per the other schemes, the server retrieves the hash of the password and the salt from the data, re-hashes the password after concatenating it with the salt, and evaluates whether the user entered media object is valid.

This scheme is beneficial when the client machine does not have sufficient processing power, for e.g. when the client is using a dumb terminal or a mobile device.

## IV. USAGE FLOW OF DIFFERENT SCHEMES

The flow for the object hash-based and the object-based schemes is illustrated in this section using screen captures. The prototype allows the user to login with either of the schemes after initial registration. We have deployed the prototype on a live web server available at [3].

*1) Sign-Up:* The client types in a username and selects a file as their password (see Fig. 4). In our prototype, both the object hash and object-based schemes share the same sign-up screen. The object file is transmitted to the server where it is processed and after multiple hashes it is stored in the database.

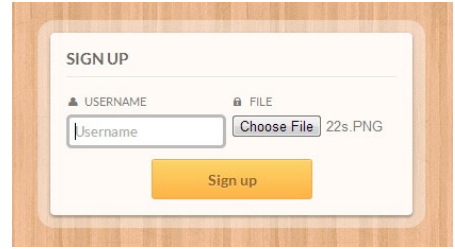

Fig. 4. Sign-up form.

*2) Login using Object:* The user logs in using the object-based scheme by simply typing in their username and selecting their object file (see Fig. 5). The object file is transmitted to the server where its hash is computed, a salt is applied and the resultant string is rehashed to be compared to the string stored in the database. The user gains access to the system if the hashes match.

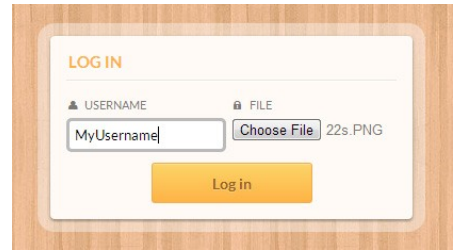

Fig. 5. Login with object-based scheme.

*3) Login using Object Hash:* Logging in using the object hash-based scheme involves a two-step procedure. The user first computes the hash of the object by selecting their password object (see Fig. 6). The hash is computed on the client system using JavaScript and the resultant hash is produced in the text-box. If the user already has the object hash stored away somewhere and opts to copy from there, they may skip this screen and proceed to the login screen.

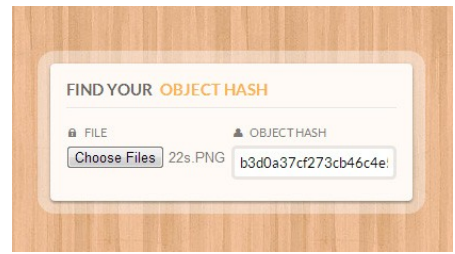

Fig. 6. Compute object hash.

Once the object hash is computed, it is copied to the clipboard and entered into the login screen (see Fig. 7). The user can also keep a copy of the object hash on another medium in case their password object are unavailable at any point in time.

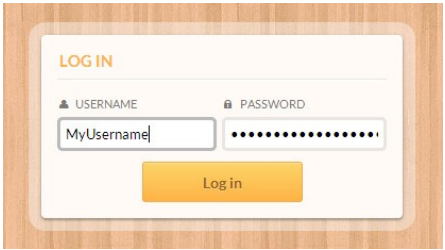

Fig. 7. Login with object hash-based scheme.

## V. EVALUATIONS

This section presents the performance evaluation of user authentication using prototype implementations of three different password schemes. Our evaluations focus on: 1) authentication delays and 2) system performance i.e. impact of using digital objects as passwords on *CPU*, memory and overall throughput.

### A. Authentication Delays

Authentication time is the time since the request is sent till the user is authenticated to the server. The time is logged when the client sends a login request and is posted to the server. Once all the computation (hashing, fetching from database) is completed and the user is authenticated, the server computes the elapsed time. In our evaluations the authentication delays are taken as an average of several measurements.

In order to measure the authentication delays the client is equipped with a 1 MB/s connection, and at the time of evaluations exhibits an average download speed of 0.76 Mbps, and an upload speed of 0.22 Mbps. The web server utilizes Intel Xeon E5620 (2.40GHz) processors whereas the client sports an Intel Core i3-2120 (3.30 GHz) processor.

*1) Text-Based Password Scheme:* In this scheme, the user is authenticated to the server on average in 1.4914 seconds. This includes network transmission time, time to retrieve the user record from the database, concatenation of salt and application of the hash function. The server computes the *SHA-256* hash (on the password concatenated with the 128-bit salt) in 0.001 seconds.

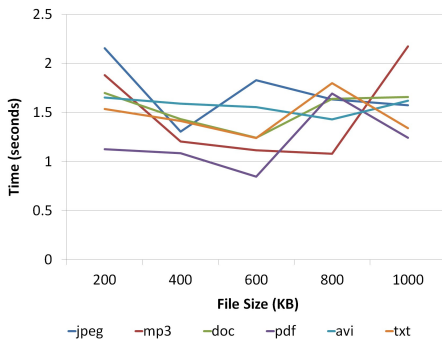

Fig. 8. Authentication time (sec) using object hash-based scheme.

*2) Object Hash-Based Password Scheme:* Performance of the object hash-based scheme is similar to that of the text-based scheme. On average, the time to authenticate the user is 1.4937 seconds (across all file formats and file sizes within our test range) against the 1.4914 seconds using the text-based scheme (see Fig. 8).

A direct relationship between file types and authentication time cannot be established. Objects are hashed in approximately the same time irrespective of their file types. The client takes only marginally longer to compute hashes of larger objects (file size greater than 1000 KB).

The time to compute the additional object hash on the client is very small as compared to the overall time to authenticate the user. It is important to note here that the time to authenticate is in the order of seconds whereas the time to compute the hash is in the order of milliseconds.

*3) Object-Based Password Scheme:* The time to authenticate the user using the object-based scheme is dominated by the object upload time. Increasing file sizes directly impacts authentication times (see Fig. 9). This is due to the increased network transmission times. The time to compute the object hash on the server is negligible. File types once again prove not to influence authentication or hashing times.

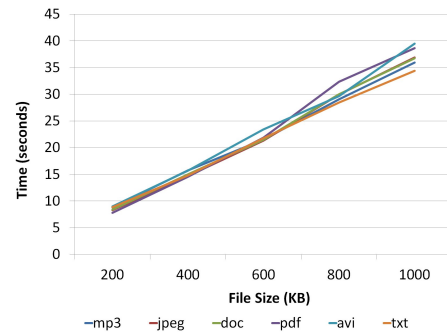

Fig. 9. Authentication time (sec) using object-based scheme.

### B. System Performance

To analyze the system performance of user authentication with text, object-hash and object based password schemes, we augment a local server with the three passwords schemes and measure their impact. We setup a server locally on 2.33 GHz Core2Duo machine with 2GB of *RAM*. In order to measure the impact of *CPU* and memory the server is bombarded with 1,200 login requests for a period of 10 minutes (2 requests per second). The experiment is repeated for each of the password schemes. The requests are produced by *JMeter* [1] deployed on another system on the *LAN*.

*It is pertinent to note that recordings of the object hash-based schemes are also valid for the text-based scheme since their functioning on the server-side is identical.*

*1) Impact on CPU Usage:* To isolate the impact on *CPU* usage we monitor only the Apache process running on the

server machine. Isolating the process also ensures *CPU* usage is 0% in idle state. *CPU* consumption on average for the object-based scheme is 6.14% (see Fig. 10). The object hash-based scheme on the other hand only consumes (on average) 0.69% of the *CPU* for the same experiment. The higher *CPU* usage for the object-based scheme can be attributed to the fact that all object hashes are computed on the server whereas on the object hash-based scheme the responsibility is delegated and distributed among the clients. To interpret these results it can be stated that usage of digital objects as password does not have a significant impact on the *CPU* load.

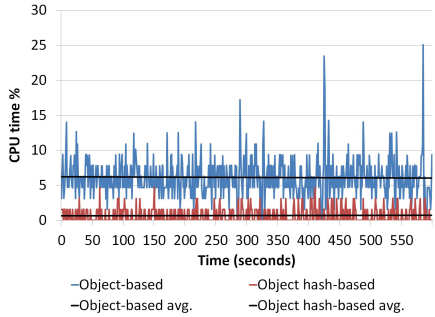

Fig. 10. CPU time of Apache process for each scheme.

*2) Impact on Memory:* Memory usage is identical for much of the 10 minute period for both object hash and object-based schemes (see Fig. 11 and 12). Memory consumption however, on average, is 2.34% higher than the idle state in either scheme with 2 gigabytes of memory available. In the end we can conclude that object based password schemes have no significant effect on *CPU* and memory resources.

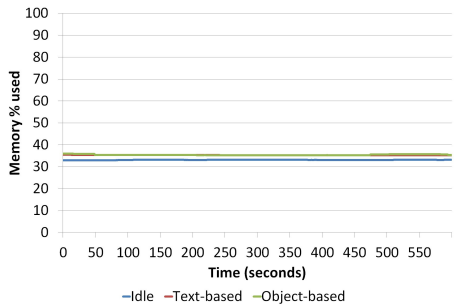

Fig. 11. Percentage of memory used for each scheme.

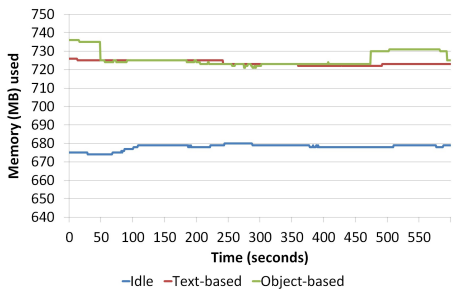

Fig. 12. Memory used in megabytes for each scheme.

*3) Throughput:* In order to measure the throughput of different password schemes, 2 clients concurrently and continuously bombarded the server with login requests (using JMeter) over a period of 10 minutes on localhost setup (2.33
GHz Core2Duo machine with 2GB of *RAM*). The number of requests divided by the time difference from the first to the last requests gives the throughput in requests per second. Even with bombardment of login requests the server is able to successfully handle 41.6 requests/second for object hash-based scheme (also valid for the text-based scheme since their functioning on the server-side is identical). For the object-based scheme the average throughput was 2.2 requests/second (with 1000 KB objects). This comparatively low throughput is because of the time required to send the objects to the server. Furthermore, since all the computation is handled by the server, the throughput very gradually keeps decreasing with time and ends up averaging to 2.2 requests/second over 10 minutes.

## VI. DISCUSSION

The three password schemes discussed previously (text-based, object hash and object-based) offer certain benefits over one another, a discussion on which is produced here.

### A. Performance

Object hash and object-based password schemes exhibit computational overheads in computing the hash of the media object. The computation (hashing) time is negligible however; standard laptops and desktop computers can compute hashes in the order of milliseconds and servers are already optimized to compute hashes of text-based passwords to be stored in the database.

A significant network cost is incurred in the case of object-based password schemes because the media object requires to be sent to the server for further computation. This scheme may be beneficial when the client machine does not have sufficient processing power, for e.g. when the client is using a dumb terminal or a mobile device.

### B. Ease of Use

Text-based password schemes offer familiarity to users, they however are not the easiest to use. Password construction at the time of signing up for a service is a cumbersome task because of various restrictions placed upon the length of the password and because of restrictions that specify what a password must (or must not) contain. Strong text-based passwords are also difficult to remember, even more so when there are multiple passwords to remember.

Object hash and object based password schemes on the other hand may not offer familiarity, but constitute a much simpler sign up process. Passwords here require selection, whereas text-based passwords require construction. It is also much simpler to explain to the user what a strong password is in terms of media objects such as photographs, than it is in terms of text. Media objects also minimize the users memory load, as they only require semantic clues to be remembered which is not usually the case for strong text-based passwords.

## C. Sharing

It is not generally recommended to share the passwords with anyone, but this is nevertheless an important feature of any password scheme. If two users have a private but pre-shared media library, such as a library of photographs, they may share a media object by sharing descriptions or cues and not the actual object to be used in the object hash or object-based scheme. The password may also be shared over a non-digital medium by first converting the media object into its text-based representation using the object hash framework.

## D. Backing-Up

Text-based passwords may be noted down on a piece of paper by the user in case they may forget what password they have set. The object hash-based scheme allows for a similar possibility by allowing the user to generate a text-based representation of their chosen media object. Media objects may also be stored on backup devices. It is also very likely that media objects, unlike text-based passwords, are already backed up. Consider for e.g. a personal family vacation album that is already backed up on a *CD-ROM*.

## E. Security

Object hash and object-based passwords are resistant to offline dictionary attacks. Considering users select a media object that is not accessible to others (users and attackers alike), attackers will find it impossible to come up with generalized look up tables. On the other hand, commonly used text-based passwords are feasible to compute [13] and expansive dictionaries containing such passwords already exist [17].

## VII. CONCLUSION

The object hash and object-based password schemes offer significant security benefits over text-based password schemes if they are used correctly. The choice between object hash and object based password schemes is a fairly simple one - the object hash framework is more efficient in performance (similar to text-based schemes) but requires computation to be performed on the client. If that computational power is not available, then the object based password scheme may be the way to go even though it will require significantly higher network resources and marginally higher server resources.

The object hash and object-based frameworks may also be used interchangeably or in concert with one another. Mobile devices could load the object-based framework whereas more powerful systems (personal computers) could utilize the object hash-based framework. Using the two interchangeably will also allow the user to obtain the text representation of the media object when operating on either framework.

It was determined through performance evaluations that media objects of differing file types (mp3, jpeg, doc, pdf, avi, txt) are hashed in virtually the same time, i.e. hashing times are independent of files types. Increasing file size does however marginally impact hashing times. This impact is marginal in the sense that it is dwarfed by the overall authentication time from the client to the server.